 \documentclass[10pt,onecolumn]{IEEEtran}
\usepackage[utf8]{inputenc}
\usepackage{amsmath}
\usepackage{mathtools}
\usepackage{amsfonts}
\usepackage{amssymb}
\usepackage{amsthm}
\usepackage{subfigure}
\usepackage{cite}
\usepackage{calc}
\usepackage{color}
\usepackage{epsfig}
\usepackage{setspace}
\usepackage{pstricks}
\usepackage{cancel}
\usepackage{multirow}
\usepackage{mathrsfs}
\usepackage{algorithmicx}
\usepackage{algpseudocode}
\usepackage{algorithm}
\usepackage{mathrsfs}

\newtheorem{defn}{Definition}
\newtheorem{thm}{{\cal T}heorem}[section]
\newtheorem{cor}[thm]{Corollary}
\newtheorem{prop}{Proposition}
\newtheorem{lem}[thm]{Lemma}
\newtheorem{conj}[thm]{Conjecture}
\newtheorem{constr}[thm]{Construction}
\newtheorem{note}{Remark}
\newtheorem{claim}{Claim}

\newcommand{\bit}{\begin{itemize}}
\newcommand{\eit}{\end{itemize}}
\newcommand{\bcor}{\begin{cor}}
\newcommand{\ecor}{\end{cor}}
\newcommand{\beq}{\begin{equation}}
\newcommand{\eeq}{\end{equation}}
\newcommand{\beqn}{\begin{equation}}
\newcommand{\eeqn}{\end{equation}}
\newcommand{\bea}{\begin{eqnarray}}
\newcommand{\eea}{\end{eqnarray}}
\newcommand{\bean}{\begin{eqnarray*}}
\newcommand{\eean}{\end{eqnarray*}}
\newcommand{\ben}{\begin{enumerate}}
\newcommand{\een}{\end{enumerate}}
\newcommand{\bdefn}{\begin{defn}}
\newcommand{\edefn}{\end{defn}}
\newcommand{\bnote}{\begin{note}}
\newcommand{\enote}{\end{note}}
\newcommand{\bprop}{\begin{prop}}
\newcommand{\eprop}{\end{prop}}
\newcommand{\blem}{\begin{lem}}
\newcommand{\elem}{\end{lem}}
\newcommand{\bthm}{\begin{thm}}
\newcommand{\ethm}{\end{thm}}
\newcommand{\bconj}{\begin{conj}}
\newcommand{\econj}{\end{conj}}
\newcommand{\bconstr}{\begin{constr}}
\newcommand{\econstr}{\end{constr}}
\newcommand{\bpf}{\begin{proof}}
\newcommand{\epf}{\end{proof}}

\title{On Epsilon-MSCR Codes for Two Erasures}

\author{
	\IEEEauthorblockN{Bh. Rekha Devi, V. Lalitha \\}
	\IEEEauthorblockA{SPCRC, International Institute of Information Technology, Hyderabad, India \\}
	Email: rekha.devi@research.iiit.ac.in, lalitha.v@iiit.ac.in}

\begin{document}

\maketitle

\begin{abstract}

Cooperative regenerating codes are regenerating codes designed to tradeoff storage for repair bandwidth in case of multiple node failures. Minimum storage cooperative regenerating (MSCR) codes are a class of cooperative regenerating codes which achieve the minimum storage point of the tradeoff. Recently, these codes have been constructed for all possible parameters $(n,k,d,h)$, where $h$ erasures are repaired by contacting any $d$ surviving nodes. However, these constructions have very large sub-packetization. $\epsilon$-MSR codes are a class of codes introduced to tradeoff subpacketization level for a slight increase in the repair bandwidth for the case of single node failures. We introduce the framework of $\epsilon$-MSCR codes which allow for a similar tradeoff for the case of multiple node failures. We present a construction of $\epsilon$-MSCR codes, which can recover from two node failures, by concatenating a class of MSCR codes  and scalar linear codes. We give a repair procedure to repair the $\epsilon$-MSCR codes in the event of two node failures and calculate the repair bandwidth for the same. We characterize the increase in repair bandwidth incurred by the method in comparison with the optimal repair bandwidth given by the cut-set bound. Finally, we show the subpacketization level of $\epsilon$-MSCR codes scales logarithmically in the number of nodes.

\end{abstract}

\section{Introduction}

In an $(n,k,d, l)$ regenerating code \cite{dimakis2010network}, a file comprised of $B$ symbols from a finite field $\mathbb{F}_q$ is encoded into a set of $nl$ code symbols and they are stored across $n$ nodes in the network with each node storing $l$ code symbols. The parameter $l$ is called the sub-packetization level of the code. A data collector can download the data by connecting to any $k$ nodes. In the event of node failure, node repair is carried out by having the replacement node connect to any $d$ nodes and download $\beta \leq l$ symbols from each node. 
A cut-set bound on the number of symbols downloaded from each of the $d$ nodes for the repair of a single node was given in \cite{dimakis2010network} as 
\begin{eqnarray}
RB \geq \frac{l}{d-k+1}.
\end{eqnarray}
The codes which achieve the above cut-set bound with equality are termed as minimum storage regenerating (MSR) codes. Explicit constructions of MSR codes for $d \geq 2k-2$ are given in \cite{rashmi2011optimal} using the product-matrix framework. Employing Hadamard designs, MSR codes which achieve high-rate for two parity nodes were presented in \cite{papailiopoulos2013repair}. In \cite{cadambe2013asymptotic}, existence of MSR codes was shown for all parameters $(n,k,d)$ as $l \rightarrow \infty$. Explicit construction of zigzag codes (high-rate systematic repair MSR codes) were given in \cite{tamo2013zigzag} for $d=n-1$. However, these codes require a sub-packetization that is exponential in $k$. 
Explicit construction of MSR codes for all parameters $(n,k,d)$ with $l=(d-k+1)^{n}$ were given in \cite{ye2017explicit}. The sub-packetization level $l$  was improved to $r^{\frac{n}{r}}$ ($r=n-k$) for $d=n-1$ by the same authors in \cite{ye2017explicitoptimal}. With the help of coupled-layer construction, MSR code with parameters $(n=qt,k=q(t-1),d=n-1,l=q^{t}=r^{\frac{n}{r}})$ was presented in \cite{sasidharan2016explicit} for auxilary parameters $q \geq 2$, $t \geq 2$. A lower bound on sub-packetization level $l$ for fixed $k$ and $n$ has been discussed in \cite{goparaju2014improved}.

\subsection{$\epsilon$-MSR Codes}

These are codes which tradeoff subpacketization for slight increase in repair bandwidth and are obtained by the concatenation of a MSR code with a scalar linear code having large minimum distance. For the repair of any failed node, the amount of information downloaded from each helper node is at-most $(1+\epsilon)$ times that of the optimal for some $\epsilon>0$. For constant $r=n-k$, the subpacketization level $l$ scales logarithmically with the number of nodes. $\epsilon$-MSR codes are known for two cases (i) All the remaining nodes are contacted for repairing the failed node \cite{rawat2018mds}. (ii) Only a subset of the remaining nodes are contacted for repair \cite{guruswami2018epsilon}. In this case, a failed node can be repaired by contacting a set of $P'$ compulsory nodes and $P-P'$ arbitrary nodes.

\subsection{Repair of Multiple Erasures}

\noindent {\bf Cooperative Repair:} In cooperative repair for $h$ node failures, node repair is carried out in two rounds. In the first round, each of the replacement node connect to any $d$ nodes of the $n-h$ surviving nodes and download $\beta_1 \leq \alpha$ symbols from each node. In the second round, the replacement nodes exchange data among themselves. Every replacement node downloads $\beta_2$ symbols from every other replacement node. Hence, the repair bandwidth per replacement node is given by
\begin{equation*}
\gamma = d \beta_1 + (h-1) \beta_2.
\end{equation*}
The cut-set bound on repair bandwidth  for $h$ failed nodes under cooperative repair (\cite{shum2013cooperative}) is given by
\begin{equation}
\label{eqn1}
RB \geq \frac{h(h+d-1)l}{h+d-k}.
\end{equation}
The codes which achieve the above cut-set bound with equality are termed as minimum storage cooperative regenerating (MSCR) codes. Constructions of MSCR codes for $d=k$ were proposed in \cite{shum2016cooperative}. Constructions of MSCR codes for $d \geq 2k-3$ which can repair two erasures were presented in \cite{li2014cooperative}. For general values of $h,d$, MSCR codes were given in \cite{ye2018cooperative} which achieve optimal repair bandwidth. These codes require large sub-packetization level $l \approx (d-k+1)^{h\binom{n}{h}}$. In a recent work, \cite{zhang2018scalar}, explicit constructions of scalar MSCR codes ($\beta = 1$) for $d \geq max\{2k-1-h,k\}$ and $l=d-k+h$ were constructed using invariant repair spaces.

\noindent {\bf Centralized Repair:} In centralized repair of $h$ node failures,  a central node recovers the contents of all the failed nodes by contacting a set of $d$ helper nodes and downloading $\beta$ symbols from each of them. The cut-set bound on repair bandwidth for $h$ failed nodes under centralized repair (\cite{cadambe2013asymptotic}) is given by
\begin{equation}
RB \geq \frac{hdl}{h+d-k}. \\
\end{equation}
The codes which achieve the above cut-set bound with equality are termed as minimum storage multi-repair (MSMR) codes.
The constructions of MSMR codes based on product-matrix codes and interference-alignment based codes have been investigated in \cite{zorgui2017centralized}. 

\subsection{Our Contributions}

In this paper, we consider the problem of constructing near-optimal minimum storage cooperative regenerating codes for two erasures. 
\begin{itemize}
\item We introduce $\epsilon$-MSCR code framework. 
\item We give a construction of  $\epsilon$-MSCR codes which are obtained by concatenating an MSCR code (which can repair two erasures) and a scalar linear code. A quick review of MSCR codes with specific parameters is given in Section \ref{sec:review_mscr} and the construction of  $\epsilon$-MSCR codes itself is presented in Section \ref{sec:emscr_repair}.
\item  We present a method to repair the code under two erasures (Section \ref{sec:emscr_detailed}) and calculate the repair bandwidth incurred by the method (Section \ref{sec:emscr_rb}).
\item We characterize the $\epsilon$ resulting from the repair method under consideration (Section \ref{sec:emscr_rb}). It turns out that $\epsilon$ is a finite value and cannot be made arbitrarily small. 
\item We also show that these codes have sub-packetization level logarithmic in the number of nodes (Section \ref{sec:emscr_rb}).
\end{itemize}


\section{Review of MSCR Codes} \label{sec:review_mscr}

In this section, we will introduce vector MDS codes and note that MSCR codes are vector MDS codes. Subsequently, we will present an $(n,k,d = k+1,h = 2,l = 3^{\binom{n}{2}})$ MSCR code which can recover from $2$ erasures by contacting $k+1$ nodes. 
\noindent {\bf Vector MDS Codes:} A linear code $\mathcal{C}_{vec}$ is said to be a vector code with parameters  $(n,K_S,d_{\min}, l)$ if $n l$ code symbols are grouped into $n$ vector-code symbols and each vector-code symbol is of size $l$ over $\mathbb{B}$. The minimum distance $d_{\min}$ of $\mathcal{C}_{vec}$ is the minimum number of vector-code symbols in which any two codewords differ or equivalently minimum weight of any non-zero codeword.  The $n l$ code symbols themselves form a $[nl, K_S]$ scalar code of dimension $K_S$ over $\mathbb{B}$. 
A vector code can be described in terms of its parity check matrix $H = [H_1, H_2 \ldots, H_n]$ as follows:
\begin{equation}
\sum_{i = 1}^n H_i \textbf{c}^i = 0,
\end{equation}
where $H_i, 1 \leq i \leq n$ is a matrix of size $(nl -K_S) \times l$ and $\textbf{c}^i$ is the vector code symbol corresponding to node $i$. A vector code is said be MDS code is $l \mid K_S$ and $d_{\min} = n - \frac{K_S}{l} + 1$. Note that MSCR codes are vector MDS codes. 

\noindent{\bf MSCR Codes:} We describe a $(n,k,d = k+1,h = 2,l = 3^{\binom{n}{2}})$ MSCR code $\mathcal{C}^Y$ over a finite field $\mathbb{B}$ via its parity check matrix as follows:
\begin{equation}
H = \begin{bmatrix} I  & I & \ldots & I \\ H_1 & H_2 & \ldots & H_n \\ \vdots & \vdots & \vdots & \vdots \\ H_1^{r-1} & H_2^{r-1} & \ldots & H_n^{r-1}   \end{bmatrix},
\end{equation}
where $H_i$ is an $l \times l$ matrix, $r = n-k$. To define $H_i$, we need to consider the $l$ scalar code-symbols in a vector code-symbol as indexed by an $m = \binom{n}{2}$-length $3$-ary representation $(b_m, b_{m-1}, \ldots, b_1)$ and $b = b_1 + 3 b_2  + \ldots + b_m 3^{m-1}$. $H_i$ is a diagonal matrix, whose $(b+1)^{\text{th}}$ diagonal entry is given by $\lambda_{i,f(i,b)} \in \mathbb{B}$. In the following, we define the function $f$.
\begin{itemize}
\item Consider a function $g(i_1,i_2) = {\binom{i_2 -1}{2}} + i_1,  1 \leq i_1 < i_2 \leq n$. $g$ maps every pair $(i_1,i_2)$ to a unique number in $\{1, 2, \ldots, {\binom{n}{2}} \}$.
\item Let $P_f = | \{ j : (1 \leq j \leq i-1) \& (b_{g(j,i)} = 2) \} | + | \{ j : (i+1 \leq j \leq n) \& (b_{g(i,j)} = 1) \} |$. 
\begin{equation}
f(i,b) = \begin{cases} 0, & \text{if } P_f \text{ is even} \\ 1, & \text{if } P_f \text{ is odd}. \end{cases}
\end{equation} 
\end{itemize}
It is required that $\{ \lambda_{1,0}, \lambda_{1,1}, \lambda_{2,0}, \lambda_{2,1}, \ldots, \lambda_{n,0}, \lambda_{n,1} \}$ are all distinct elements in $\mathbb{B}$. The following claim gives the structure of the function $f$ which will be used in the later parts of the paper.
\begin{claim} \label{cl:propf}
Consider a pair $(i_1, i_2)$ such that $1 \leq i_1 < i_2 \leq n$. We denote $g(i_1,i_2) = g_{12}$. Also, let $b(i,u)$ be defined as follows:
\begin{equation}
b(i,u) = (b_{m},b_{m-1},...,b_{i+1},u,b_{i-1},...,b_{1}).
\end{equation}
 Then, based on the definition of function $f$ above,  we have 
 \begin{eqnarray*}
&& f(i,b(g_{12},0)) =  f(i,b(g_{12},1)) =  f(i,b(g_{12},2)), i \not \in \{i_1,i_2\} \\
 && f(i_1,b(g_{12},0)) =  f(i_1,b(g_{12},2)) \not = f(i_1,b(g_{12},1)), \\
&& f(i_2,b(g_{12},0)) =  f(i_2,b(g_{12},1)) \not = f(i_2,b(g_{12},2)).
 \end{eqnarray*}
\end{claim}

\section{$\epsilon$-MSCR Codes for Two Erasures} \label{sec:emscr_repair}

\begin{defn}
Consider a vector MDS code $\mathcal{C}$ with parameters $(n, kl , d_{\min} = n-k+1, l)$. For any $\epsilon > 0$, we say that $\mathcal{C}$ is an $\epsilon$-MSCR code for $h$ erasures, if any $h$ erasures can be repaired by contacting $d$ of the remaining nodes with a total repair bandwidth which is at most $(1+\epsilon) \frac{h(h+d-1)l}{(h+d-k)}$.
\end{defn}

In the following, we give a construction of $\epsilon$-MSCR code which can be recovered from two erasures.

\begin{constr} \label{constr:emscr}
An $\epsilon$-MSCR code is a vector MDS code obtained by concatenating an MSCR code which can recover from two node failures and a scalar linear code. We will first describe the parameters of the constituent codes and then give the method of concatenation. 

\noindent {\bf MSCR Code:} The first constituent code is an MSCR code with parameters $(n,k,d = k+1,h = 2,l = 3^{\binom{n}{2}})$ described in Section \ref{sec:review_mscr}. We consider the case when $r = n-k \geq 5$. $\{ \lambda_{1,0}, \lambda_{1,1}, \lambda_{2,0}, \lambda_{2,1}, \ldots, \lambda_{n,0}, \lambda_{n,1} \} \in \mathbb{B}_0 \subset \mathbb{B}$, where $\mathbb{B}_0$ is a multiplicative subgroup of $\mathbb{B}\setminus \{0\}$.

\noindent {\bf Scalar Code:} The second constituent code is a scalar linear code over $\mathbb{F}_q, r = n-k < q \leq n$ with length $N$, dimension $K$ and minimum distance $D = \delta N$, $0 < \delta < 1$. The number of codewords in the scalar code is assumed to be $M$. We will denote this code by $\mathcal{C}^S$.

\noindent {\bf $\epsilon$-MSCR Code:} Given the above two codes, $\epsilon$-MSCR code $\mathcal{C}$ is a vector MDS code with the parameters $(M, K_S = (M-r) N l, d_{\min} = r+1, L = Nl)$. The number of nodes in the $\epsilon$-MSCR code equals the number of codewords in the scalar code $\mathcal{C}^S$. The nodes themselves are indexed by the codewords of the scalar code $\mathcal{C}^S$. Let $\textbf{a}_i = (a_{i,1}, a_{i,2}, \ldots a_{i,N})$ denote the $i^{\text{th}}$ codeword of the scalar code. The parity check equation satisfied by the $\epsilon$-MSCR code $\mathcal{C}$ is given by
\begin{equation}
\sum_{i=1}^M \mathcal{H}_i \textbf{c}^i = 0, \text{where}
\end{equation}
\begin{equation} \label{eq:pc_emscr}
\mathcal{H}_i = \begin{bmatrix} \text{Diag}(I, I, \ldots, I) \\ \sigma_i \text{Diag}(H_{a_{i,1}}, H_{a_{i, 2}}, \ldots, H_{a_{i, N}}) \\ \vdots \\ \sigma_i^{r-1} \text{Diag}(H_{a_{i,1}}^{r-1}, H_{a_{i, 2}}^{r-1}, \ldots, H_{a_{i, N}}^{r-1}) \end{bmatrix}. 
\end{equation}
$\{ \sigma_i, 1 \leq i \leq M \}$ are picked such that each $\sigma_i$ belongs to a distinct coset of $\mathbb{B}_0$ in $\mathbb{B}\setminus\{0\}$.
\end{constr}

\begin{thm}
The code $\mathcal{C}$ given in Construction \ref{constr:emscr} is a vector MDS code.
\end{thm}
\begin{IEEEproof}
The code is defined by its $rNl \times MNl$ parity check matrix $\mathcal{H}$. Any $rNl \times rNl$ sub-matrix of $\mathcal{H}$ should be full-rank for $\epsilon$-MSCR to satisfy the MDS property. Each thick column($Nl \times 1$) of $\mathcal{H}$ is indexed by a codeword of $\mathcal{C}^{S}$ by the construction of the code. Let the $r$ distinct codewords of $\mathcal{C}^{S}$ indexing the $rNl$ columns of $\mathcal{H}$ be $\mathcal{A}=\{a_{1},a_{2},..,a_{r}\}$.
Then the $rNl \times rNl$ parity-check matrix corresponding to these codewords is given by
\[\mathcal{H}_{\mathcal{A}}=\begin{bmatrix}
\mathcal{H}_{a_{1}}&\mathcal{H}_{a_{2}}&\cdots&\mathcal{H}_{a_{r}}
\end{bmatrix}\]
\[=\begin{bmatrix}
Diag(I,\cdots,I)&\cdots&Diag(I,\cdots,I)\\
\sigma_{1}Diag(H_{a_{1,1}},\cdots,H_{a_{1,N}})&\cdots&\sigma_{r}Diag(H_{a_{r,1}},\cdots,H_{a_{r,N}})\\
\vdots&\vdots&\vdots\\
\sigma_{1}^{r-1}Diag(H_{a_{1,1}}^{r-1},\cdots,H_{a_{1,N}}^{r-1})&\cdots&\sigma_{r}^{r-1}Diag(H_{a_{r,1}}^{r-1},\cdots,H_{a_{r,N}}^{r-1})\\
\end{bmatrix}.\]
Because of the block diagonal structure of the $Nl \times Nl$ sub-matrices in the above equation, we need to only show that the following matrix is full-rank, for all $i \in [N]$.  
\[U_{\mathcal{A},i}=\begin{bmatrix}
I&\cdots&I\\
\sigma_{1}H_{a_{1,i}}&\cdots&\sigma_{r}H_{a_{r,i}}\\
\vdots&\vdots&\vdots\\
\sigma_{1}^{r-1}H_{a_{1,i}}^{r-1}&\cdots&\sigma_{r}^{r-1}H_{a_{r,i}}^{r-1}
\end{bmatrix}\]
\[=\begin{bmatrix}
Diag(1,\cdots,1)&\cdots&Diag(1,\cdots,1)\\
\sigma_{1}Diag(\lambda_{a_{1,i},f(a_{1,i},0)},\cdots,\lambda_{a_{1,i},f(a_{1,i},l-1)})&\cdots&\sigma_{r}Diag(\lambda_{a_{r,i},f(a_{r,i},0)},\cdots,\lambda_{a_{r,i},f(a_{r,i},l-1)})\\
\vdots&\vdots&\vdots\\
\sigma_{1}^{r-1}Diag(\lambda_{a_{1,i},f(a_{1,i},0)}^{r-1},\cdots,\lambda_{a_{1,i},f(a_{1,i},l-1)}^{r-1})&\cdots&\sigma_{r}^{r-1}Diag(\lambda_{a_{r,i},f(a_{r,i},0)}^{r-1},\cdots,\lambda_{a_{r,i},f(a_{r,i},l-1)}^{r-1})
\end{bmatrix}\]
Re-arranging the rows and columns, we get
\[=\begin{bmatrix}
\begin{bmatrix}
1&\cdots&1\\
\sigma_{1}\lambda_{a_{1,i},f(a_{1,i},0)}&\cdots&\sigma_{r}\lambda_{a_{r,i},f(a_{r,i},0)}\\
\vdots&\vdots&\vdots\\
\sigma_{1}^{r-1}\lambda_{a_{1,i},f(a_{1,i},0)}^{r-1}&\cdots&\sigma_{r}^{r-1}\lambda_{a_{r,i},f(a_{r,i},0)}^{r-1}\\
\end{bmatrix}&\cdots&0\\
&\ddots\\
0&\cdots&\begin{bmatrix}
1&\cdots&1\\
\sigma_{1}\lambda_{a_{1,i},f(a_{1,i},l-1)}&\cdots&\sigma_{r}\lambda_{a_{r,i},f(a_{r,i},l-1)}\\
\vdots&\vdots&\vdots\\
\sigma_{1}^{r-1}\lambda_{a_{1,i},f(a_{1,i},l-1)}^{r-1}&\cdots&\sigma_{r}^{r-1}\lambda_{a_{r,i},f(a_{r,i},l-1)}^{r-1}\\
\end{bmatrix}
\end{bmatrix}.\]
$U_{\mathcal{A},i}$ is a block-diagonal matrix where each diagonal block is a Vandermonde matrix. Hence it is a full-rank matrix, $\forall i \in [N]$ completing the proof.\\   
\end{IEEEproof}

\begin{note}
It is clear that using $(n,k,d)$ MSR codes, simultaneous repair of multiple erasures can be performed. However, it is not possible for the case of $\epsilon$-MSR codes. This is because for repair of $\epsilon$-MSR codes when all the remaining nodes are not contacted, the code has to satisfy  $(P, P')$ repair property and there is a set of $P'$ compulsory nodes which have to be contacted. If the second erasure is from one of the compulsory nodes, then we cannot recover from the two erasures.
\end{note}

\section{Repair of $\epsilon$-MSCR Codes for Two Erasures} \label{sec:emscr_detailed}

In this section, we will describe the repair of $\epsilon$-MSCR codes for two erasures. We would like to note that as in the case of $\epsilon$-MSR codes, an $\epsilon$-MSCR code is said to have $(P, P')$ repair property if for repairing $h=2$ erasures, $P$ nodes are contacted, $P'$ of which are compulsory nodes and the remaining $P - P'$ nodes are arbitrary.

Suppose that the nodes $c^{1}$ and $c^{2}$ indexed by codewords $a_{1},a_{2} \in \mathcal{C}^{S}$ have failed. 
\eqref{eq:pc_emscr} represents the parity check column $\mathcal{H}_{i}$ of $\mathcal{C}$ corresponding to any codeword $a_{i} \in \mathcal{C}^{S}$. We give the repair procedure for repairing  $c^{1}_{1}$ and $c^{2}_{1}$. The same procedure can be applied for repairing all $c^{1}_{j}, c^{2}_{j}, j \in [2,N]$, since all parity check columns are block diagonal matrices. Consider the columns of $\mathcal{H}$ corresponding to $a_{i,1}, i\in[M]$

\[\begin{bmatrix}
I&I&\cdots &I\\
\sigma_{1}H_{a_{1,1}}&\sigma_{2}H_{a_{2,1}}&\cdots&\sigma_{M}H_{a_{M,1}}\\
\vdots &\vdots&\cdots&\vdots\\
\sigma_{1}^{r-1}H_{a_{1,1}}^{r-1}&\sigma_{2}^{r-1}H_{a_{2,1}}^{r-1}&\cdots&\sigma_{M}^{r-1}H_{a_{M,1}}^{r-1}\\
\end{bmatrix}.\]
The parity check equation corresponding to $a_{i,1}, i \in [M]$ is given by
\begin{eqnarray}
\sum_{i=1}^{M} \sigma_{i}^{t} H_{a_{i,1}}^{t} c_{1}^{i}=0,  \quad t \in [0,r-1].
\end{eqnarray}

\begin{note}
We would like to note here that the repair procedure is different for the case when $a_{1,1} \neq a_{2,1}$ and for the case when $a_{1,1} = a_{2,1}$. This is because whenever $a_{1,1} \neq a_{2,1}$, then based on the construction of MSCR code, the function $g(a_{1,1}, a_{2,1})$ denoted as $g_{12}$ is well defined if $a_{1,1} < a_{2,1}$. Otherwise we use the function $g(a_{2,1}, a_{1,1})$ denoted as $g_{21}$ and hence the repair is performed based on partitioning the indices according to this function. However, when $a_{1,1} = a_{2,1}$, then the function $g(a_{1,1}, a_{2,1})$ is not defined and for $a_{3,1} \neq (a_{1,1} = a_{2,1})$, we perform the repair based on partitioning the indices with respect to $g(a_{1,1}, a_{3,1})$ denoted as $g_{13}$ if $a_{1,1} < a_{3,1}$. Otherwise we use the function $g(a_{3,1}, a_{1,1})$ denoted as $g_{31}$.
\end{note}

\subsection{Case 1: $a_{1,1} \neq a_{2,1}$}
We construct three disjoint sets $Q, V, \Gamma$ based on the codewords in $C^S$ as follows:
\begin{eqnarray*}
Q & = & \{i: a_{i,1}  =  a_{1,1} ,i\in[3,M]\}, \\
V & = & \{i: a_{i,1}  =  a_{2,1}, i \in[3,M]\}, \\
 \Gamma & =  & \{i: a_{i,1} \not = a_{1,1}, \ \   a_{i,1} \not =  a_{2,1} , i\in[3,M] \} . 
\end{eqnarray*}
\begin{eqnarray*}
 \sigma_{1}^{t}.H_{a_{1,1}}^{t}.c_{1}^{1}+ \sigma_{2}^{t}.H_{a_{2,1}}^{t}.c_{1}^{2} 
 + \sum_{q_{i}\in Q}\sigma_{q_{i}}^{t}.H_{a_{q_{i},1}}^{t}.c_{1}^{q_{i}} +\sum_{v_{i}\in V}\sigma_{v_{i}}^{t}.H_{a_{v_{i},1}}^{t}.c_{1}^{v_{i}} 
 +\sum_{\gamma_{i}\in \Gamma} \sigma_{\gamma_{i}}^{t}.H_{a_{\gamma_{i},1}}^{t}.c_{1}^{\gamma_{i}} = 0 , t \in [0,r-1].
\end{eqnarray*}

Considering the parity check equation corresponding to  $b(g_{12},k)$, $k \in [0,2]$ and substituting $\forall q_{i} \in Q,a_{q_{i},1}=a_{1,1}$ and $\forall v_{i} \in V,a_{v_{i},1}=a_{2,1}$, we have

 \begin{IEEEeqnarray*}{l}
 \label{eqn10}
\sigma_{1}^{t} \lambda_{a_{1,1},f(a_{1,1},b(g_{12},k))}^{t}c_{1,b(g_{12},k)}^{1}
+ \sigma_{2}^{t} \lambda_{a_{2,1},f(a_{2,1},b(g_{12},k))}^{t}c_{1,b(g_{12},k)}^{2} 
+\sum_{q_{i} \in Q} \sigma_{q_{i}}^{t} \lambda_{a_{1,1},f(a_{1,1},b(g_{12},k))}^{t}c_{1,b(g_{12},k)}^{q_{i}}\\
+ \sum_{v_{i}\in V} \sigma_{v_{i}}^{t} \lambda_{a_{2,1},f(a_{2,1},b(g_{12},k))}^{t}c_{1,b(g_{12},k)}^{v_{i}}
+ \sum_{\gamma_{i}\in \Gamma} \sigma_{\gamma_{i}}^{t} \lambda_{a_{\gamma_{i},1},f(a_{\gamma_{i},1},b(g_{12},k))}^{t}c_{1,b(g_{12},k)}^{\gamma_{i}} = 0,
\quad t\in[0,r-1],k\in[0,2].
\IEEEyesnumber \\
\end{IEEEeqnarray*}

Applying Claim \ref{cl:propf} with $i = a_{i,1}, i_1 = a_{1,1}, i_2 = a_{2,1}$, we can define the following:
 \begin{IEEEeqnarray*}{l}
 \label{eqn11}
 \lambda_{a_{i,1}}:= \lambda_{a_{i,1},f(a_{i,1},b(g_{12},0))} =  \lambda_{a_{i,1},f(a_{i,1},b(g_{12},1))} =  \lambda_{a_{i,1},f(a_{i,1},b(g_{12},2))} ,  a_{i,1} \not \in \{a_{1,1},a_{2,1}\}
 \IEEEyesnumber\\
 \label{eqn12}
\lambda_{a_{1,1},0}^{'}:= \lambda_{a_{1,1},f(a_{1,1},b(g_{12},0))} =  \lambda_{a_{1,1},f(a_{1,1},b(g_{12},2))}, \qquad
\lambda_{a_{1,1},1}^{'}:= \lambda_{a_{1,1},f(a_{1,1},b(g_{12},1))}, 
\IEEEyesnumber\\
\label{eqn13}
\lambda_{a_{2,1},0}^{'}:= \lambda_{a_{2,1},f(a_{2,1},b(g_{12},0))} =  \lambda_{a_{2,1},f(a_{2,1},b(g_{12},1))}, \qquad
\lambda_{a_{2,1},1}^{'}:= \lambda_{a_{2,1},f(a_{2,1},b(g_{12},2))}.\IEEEyesnumber 
\end{IEEEeqnarray*}
${\lambda_{a_{1,1},0}^{'},\lambda_{a_{1,1},1}^{'},\lambda_{a_{2,1},0}^{'},\lambda_{a_{2,1},1}^{'},\lambda_{a_{i,1}}},$
$a_{i,1} \not \in \{a_{1,1},a_{2,1}\}$ are all different.

Using the notation defined in (\ref{eqn11})-(\ref{eqn13}), we can write (\ref{eqn10}) for $k \in [0,1]$ 
and sum over $k \in [0,1]$ resulting in
\begin{IEEEeqnarray*}{l}
\label{eqn14}
\sigma_{1}^{t}\sum_{k=0}^{1} (\lambda_{a_{1,1},k}^{'})^{t}c_{1,b(g_{12},k)}^{1}
+ \sigma_{2}^{t} (\lambda_{a_{2,1},0}^{'})^{t}\sum_{k=0}^{1}c_{1,b(g_{12},k)}^{2}
+\sum_{q_{i} \in Q} \sigma_{q_{i}}^{t} \sum_{k=0}^{1} (\lambda_{a_{1,1},k}^{'})^{t}c_{1,b(g_{12},k)}^{q_{i}} \\
+ \sum_{v_{i}\in V} \sigma_{v_{i}}^{t} (\lambda_{a_{2,1},0}^{'})^{t}\sum_{k=0}^{1}c_{1,b(g_{12},k)}^{v_{i}}
+ \sum_{\gamma_{i}\in \Gamma} \sigma_{\gamma_{i}}^{t} (\lambda_{a_{\gamma_{i},1}})^{t}\sum_{k=0}^{1}c_{1,b(g_{12},k)}^{\gamma_{i}} = 0 , \quad t \in [0,r-1].
 \IEEEyesnumber 
\end{IEEEeqnarray*}
Let
\begin{IEEEeqnarray*}{l}
\mu_{2,1,1}^{(b)} = \sum_{k=0}^{1} c^{2}_{1,b(g_{12},k)},\quad
\mu_{v_{i},1,1}^{(b)} = \sum_{k=0}^{1} c^{v_{i}}_{1,b(g_{12},k)}, \quad
\mu_{\gamma_{i},1,1}^{(b)} = \sum_{k=0}^{1} c^{\gamma_{i}}_{1,b(g_{12},k)}. 
\end{IEEEeqnarray*}
The eq(\ref{eqn14}) is of the form
\begin{IEEEeqnarray*}{l}
\label{eqn15}
L_{1}+L_{2}+L_{3}+L_{4}+L_{5}=0, \IEEEyesnumber
\textnormal{where}
\end{IEEEeqnarray*}
\begin{IEEEeqnarray*}{l}
L_{j}=E_{L_j} F_{L_j}, j \in \{1,3\}, \:
 L_{2}=\sum_{q_{i}\in Q} E_{q_{i}}F_{q_{i}},  \\
L_{4}=E_{V}F_{V} , L_{5}=E_{R}F_{R} .\: \textnormal{Particularly}\\
E_{L_{1}}=[\sigma_{1}^{t}(\lambda_{a_{1,1},k}^{'})^{t}]_{t \in [0,r-1],k \in [0,1]}, \qquad
F_{L_{1}}=[c_{1,b(g_{12},k)}^{1}]_{k \in [0,1]},\\
E_{q_{i}}=[\sigma_{q_{i}}^{t}(\lambda_{a_{1,1},k}^{'})^{t}]_{t \in [0,r-1],k \in [0,1]}, \qquad\:
F_{q_{i}}=[c_{1,b(g_{12},k)}^{q_{i}}]_{k \in [0,1]},\\
E_{L_{3}}=[\sigma_{2}^{t}(\lambda_{a_{2,1},0}^{'})^{t}]_{t \in [0,r-1]}, \qquad\qquad\:\:\:
F_{L_{3}}=\mu_{2,1,1}^{(b)},\\
E_{V}=[\sigma_{v_{i}}^{t}(\lambda_{a_{2,1},0}^{'})^{t}]_{t \in [0,r-1],i \in [1,|V|]},\quad\:\:\:
F_{V}=[\mu_{v_{i},1,1}^{(b)}]_{i \in [1,|V|]},\\
E_{R}=[\sigma_{\gamma_{i}}^{t}(\lambda_{a_{\gamma_{i},1}})^{t}]_{t \in [0,r-1],i \in [1,|\Gamma|]}, \qquad
F_{R}=[\mu_{\gamma_{i},1,1}^{(b)}]_{i \in [1,|\Gamma|]}.
\end{IEEEeqnarray*}
We now construct a matrix $P_{1}$ and on left-multiplying (\ref{eqn15}) with it, we get
\begin{IEEEeqnarray*}{l}
P_{1}L_{1}+P_{1}L_{2}+P_{1}L_{3}+P_{1}L_{4}+P_{1}L_{5}=0. \\
p_0(x)=\Pi_{k=0}^{1}(x-\sigma_{1}\lambda_{a_{1,1},k}^{'})(x-\sigma_{2}\lambda_{a_{2,1},0}^{'}) \quad
\textnormal{and} \\
p_{i}(x)=x^{i}p_{0}(x) \quad \textnormal{for} \: i=0,1,..r-4. \\ \qquad
\end{IEEEeqnarray*}
For all $i \in [0,r-4]$, the degree of $p_{i}(x) < r$, hence
\begin{IEEEeqnarray*}{l}
p_{i}(x)=\sum_{j=0}^{r-1}p_{ij}x^{j} , \qquad i=0,1,..r-4. 
\end{IEEEeqnarray*}
The $(r-3)\times r$ matrix $P_{1}$ is defined as 
 \[P_{1} = \begin{bmatrix}
p_{0,0}&p_{0,1}&\cdots &p_{0,r-1}\\
p_{1,0}&p_{1,1}&\cdots &p_{1,r-1}\\
\vdots & \vdots & \vdots & \vdots\\
p_{r-4,0}&p_{r-4,1}&\cdots &p_{r-4,r-1}\\
\end{bmatrix}.\]\\

\[P_{1}(L_{1}+L_{3}) =P_{1}\begin{bmatrix}
E_{L_{1}}&E_{L_{3}}
\end{bmatrix}
\begin{bmatrix}
F_{L_{1}}\\
F_{L_{3}}
\end{bmatrix}\qquad\qquad\qquad\quad\]
\[=P_{1}\begin{bmatrix}
1&1&1\\
\sigma_{1}\lambda_{a_{1,1},0}^{'}&\sigma_{1}\lambda_{a_{1,1},1}^{'} &\sigma_{2}\lambda_{a_{2,1},0}^{'}\\
\vdots & \vdots & \vdots\\
\sigma_{1}^{r-1}(\lambda_{a_{1,1},0}^{'})^{r-1}&\sigma_{1}^{r-1}(\lambda_{a_{1,1},1}^{'})^{r-1} &\sigma_{2}^{r-1}(\lambda_{a_{2,1},0}^{'})^{r-1}\\
\end{bmatrix}
\begin{bmatrix}
F_{L_{1}}\\
F_{L_{3}}
\end{bmatrix}\]\\
\[=\begin{bmatrix}
p_{0}(\sigma_{1}\lambda_{a_{1,1},0}^{'})&p_{0}(\sigma_{1}\lambda_{a_{1,1},1}^{'})&p_{0}(\sigma_{2}\lambda_{a_{2,1},0}^{'})\\
\vdots  &\cdots& \vdots\\
p_{r-4}(\sigma_{1}\lambda_{a_{1,1},0}^{'})&p_{r-4}(\sigma_{1}\lambda_{a_{1,1},1}^{'})&p_{r-4}(\sigma_{2}\lambda_{a_{2,1},0}^{'})\\
\end{bmatrix}
\begin{bmatrix}
F_{L_{1}}\\
F_{L_{3}}
\end{bmatrix}
=0.\]


\[P_{1}L_{2}=P_{1}\sum _{q_{i} \in Q}
E_{q_{i}}. F_{q_{i}}
\qquad\qquad\qquad\qquad\qquad\qquad\quad\]
\[=\sum _{q_{i} \in Q}P_{1}.E_{q_{i}}. F_{q_{i}}
\qquad\qquad\qquad\qquad\qquad\]

\[=\sum _{q_{i} \in Q}\begin{bmatrix}
p_{0}(\sigma_{q_{i}}\lambda_{a_{1,1},0}^{'})&p_{0}(\sigma_{q_{i}}\lambda_{a_{1,1},1}^{'})\\
\vdots&\vdots\\
p_{r-4}(\sigma_{q_{i}}\lambda_{a_{1,1},0}^{'})&p_{r-4}(\sigma_{q_{i}}\lambda_{a_{1,1},1}^{'})\\
\end{bmatrix}
\begin{bmatrix}
c^{q_{i}}_{1,b(g_{12},0)}\\
c^{q_{i}}_{1,b(g_{12},1)}
\end{bmatrix}.\]
We can compute $P_{1}L_{2}$ at the first replacement node by downloading the symbols $\{c^{q_{i}}_{1,b(g_{12},0)},c^{q_{i}}_{1,b(g_{12},1)}\}$ of $F_{q_{i}}, \forall q_{i} \in Q$.

\[P_{1}L_{4}=P_{1}.E_{V}.F_{V}
\qquad\qquad\qquad\qquad\qquad\qquad\qquad\]

\[=\begin{bmatrix}
p_{0}(\sigma_{v_{1}}\lambda_{a_{2,1},0}^{'})&\cdots &p_{0}(\sigma_{v_{|V|}}\lambda_{a_{2,1},0}^{'})\\
\vdots & \vdots & \vdots\\
p_{r-4}(\sigma_{v_{1}}\lambda_{a_{2,1},0}^{'})&\cdots &p_{r-4}(\sigma_{v_{|V|}}\lambda_{a_{2,1},0}^{'})\\
\end{bmatrix}
\begin{bmatrix}
\mu_{v_{1},1,1}^{(b)}\\
\vdots\\
\mu_{v_{|V|,1,1}}^{(b)}
\end{bmatrix}.\]
We can compute $P_{1}L_{4}$ at the first replacement node by downloading the symbols $\{ \mu_{v_{i},1,1}^{(b)}\}$ of $F_{V}, \forall v_{i} \in V$.

\[P_{1}L_{5}=P_{1}.E_{R}.F_{R}
\qquad\qquad\qquad\qquad\qquad\qquad\qquad\]

\[=\begin{bmatrix}
p_{0}(\sigma_{\gamma_{1}}\lambda_{a_{\gamma_{1},1}})&\cdots &p_{0}(\sigma_{\gamma_{|\Gamma|}}\lambda_{a_{\gamma_{|\Gamma|},1}})\\
\vdots & \vdots & \vdots\\
p_{r-4}(\sigma_{\gamma_{1}}\lambda_{a_{\gamma_{1},1}})&\cdots &p_{r-4}(\sigma_{\gamma_{|\Gamma|}}\lambda_{a_{\gamma_{|\Gamma|},1}})\\
\end{bmatrix}
\begin{bmatrix}
\mu_{\gamma_{1},1,1}^{(b)}\\
\vdots\\
\mu_{\gamma_{|\Gamma|},1,1}^{(b)}
\end{bmatrix}.\]

$p_{0}(\sigma_{\gamma_{1}}\lambda_{a_{\gamma_{1},1}})\cdots p_{0}(\sigma_{\gamma_{|\Gamma|}}\lambda_{a_{\gamma_{|\Gamma|},1}})$ are all non-zero. $P_{1}L_{5}$ is a full-rank matrix. All $(r-3) \times (r-3)$ sub-matrices of $P_{1}L_{5}$ also are full-rank which follows from its Vandermonde like structure. Hence, we have
\begin{IEEEeqnarray*}{l}
\label{eqn16}
P_{1}L_{5} = -P_{1}L_{2}-P_{1}L_{4}.
\IEEEyesnumber
\end{IEEEeqnarray*}
By downloading any subset of size $k^{'}=(|\Gamma|-(r-3))$ from $F_{R}$ ($|\Gamma|-(r-3) > 0 $ since $q > r$ and $|\Gamma| \geq q-2$), the remaining can be recovered from (\ref{eqn16}). This is because after substituting $k^{'}$ values in (\ref{eqn16}) and rewriting, it would result in $(r-3)$ equations in $(r-3)$ variables, which can be solved.  From $(L_{1}+L_{3})=-L_{2}-L_{4}-L_{5}$, we have
\begin{equation}
\begin{bmatrix} E_{L_1} & E_{L_3} \end{bmatrix} \begin{bmatrix} F_{L_1} \\ F_{L_3} \end{bmatrix} = -L_{2}-L_{4}-L_{5}.
\end{equation}
By inverting a square submatrix of $\begin{bmatrix} E_{L_1} & E_{L_3} \end{bmatrix}$, we can recover $F_{L_{1}}$ and $F_{L_{3}}$. \\
\\
We can write an equation similar to (\ref{eqn14}) for $k \in \{0,2\}$ 
and by performing similar calculations as above, we can recover $\{c^{2}_{1,b(g_{12},0)}\} \cup \{c^{2}_{1,b(g_{12},2)}\} \cup \{\sum_{k=0,2} c^{1}_{1,b(g_{12},k)}\}$ at the second replacement node.\\

Please refer to the table for the summary of the downloads and recovery in the two rounds. After both the rounds, first replacement node recovers
$\{c^{1}_{1,b(g_{12},k)}:b_{g_{12}}=0,k \in \{0,1,2\}\} = \{c_{1,b}^{1}:b \in \{0,1,..l-1\}\}$
and second replacement node recovers
$\{c^{2}_{1,b(g_{12},k)}:b_{g_{12}}=0,k \in \{0,1,2\}\} = \{c_{1,b}^{2}:b \in \{0,1,..l-1\}\}.$
Both $c^{1}_{1}$ and $c^{2}_{1}$ are recovered.

\begin{itemize}
\item The repair bandwidth for the case $a_{1,1} \neq a_{2,1}$ is given by 
\begin{IEEEeqnarray*}{rCl's}
RB_{a_{1,1} \not = a_{2,1}} = k^{'}\frac{l}{3} + |Q|\frac{2l}{3} + |V|\frac{l}{3} +\frac{l}{3} \\
+ k^{'}\frac{l}{3} + |Q|\frac{l}{3} + |V|\frac{2l}{3} +\frac{l}{3} \\
= k^{'}(\frac{2l}{3}) + |Q|l+|V|l +\frac{2l}{3}. \IEEEyesnumber \label{eq:a1neqa2}
\end{IEEEeqnarray*}
where $k^{'}=|\Gamma|-(r-3)$ and $|\Gamma|=M-\frac{2M}{q}$.
\item Note that we need to contact all nodes in $Q,V$ compulsorily.
\end{itemize}

\subsection{Case 2: $a_{1,1}  = a_{2,1}$}
Consider a node indexed by codeword $a_3$ in $\mathcal{C}^S$ such that $\forall j \in [N]$, $a_{1,j} = a_{2,j}$, $a_{3,j} \neq (a_{1,j} = a_{2,j})$. Such a codeword exists in $\mathcal{C}^S$ if there is a codeword in $\mathcal{C}^S$ of Hamming weight $N$. (The existence of such codewords is guaranteed as we use the same scalar linear code used in \cite{guruswami2018epsilon}). 
Based on $a_3$, we build three sets $W, Y, Z$  where 
\begin{IEEEeqnarray*}{l}
W=\{i:a_{i,1}=a_{1,1}=a_{2,1},i\in[4,M]\}, \\
Y=\{i:a_{i,1}=a_{3,1},i\in[4,M]\} ,\\
Z=\{i:a_{i,1} \not= (a_{1,1}=a_{2,1}) \:\:\&\& \:\:a_{i,1} \not= a_{3,1},i\in[4,M]\} . 
\end{IEEEeqnarray*}
\begin{IEEEeqnarray*}{l}
\sigma_{1}^{t}.H_{a_{1,1}}^{t}.c_{1}^{1}
+\sigma_{2}^{t}.H_{a_{2,1}}^{t}.c_{1}^{2}
+\sum_{w_{i}\in W}\sigma_{w_{i}}^{t}.H_{a_{w_{i},1}}^{t}.c_{1}^{w_{i}}
+\sigma_{3}^{t}.H_{a_{3,1}}^{t}.c_{1}^{3}
+\sum_{y_{i}\in Y}\sigma_{y_{i}}^{t}.H_{a_{y_{i},1}}^{t}.c_{1}^{y_{i}}\\
+\sum_{z_{i}\in Z}
\sigma_{z_{i}}^{t}.H_{a_{z_{i},1}}^{t}.c_{1}^{z_{i}} = 0 , \quad t \in [0,r-1].
\IEEEyesnumber
\end{IEEEeqnarray*}

Considering the parity check equation corresponding to  $b(g_{13},k)$, $k \in [0,2]$  and substituting  
$a_{2,1} = a_{1,1}, \forall w_{i} \in W , a_{w_{i},1} = a_{1,1}$ and $\forall y_{i} \in Y , a_{y_{i},1} = a_{3,1}$, we have

 \begin{IEEEeqnarray*}{l}
 \label{eqn18}
\sigma_{1}^{t} \lambda_{a_{1,1},f(a_{1,1},b(g_{13},k))}^{t}c_{1,b(g_{13},k)}^{1}
+ \sigma_{2}^{t} \lambda_{a_{1,1},f(a_{1,1},b(g_{13},k))}^{t}c_{1,b(g_{13},k)}^{2} 
+\sum_{w_{i} \in W} \sigma_{w_{i}}^{t} \lambda_{a_{1,1},f(a_{1,1},b(g_{13},k))}^{t}c_{1,b(g_{13},k)}^{w_{i}} \\
+ \sigma_{3}^{t} \lambda_{a_{3,1},f(a_{3,1},b(g_{13},k))}^{t}c_{1,b(g_{13},k)}^{3} 
+ \sum_{y_{i}\in Y} \sigma_{y_{i}}^{t} \lambda_{a_{3,1},f(a_{3,1},b(g_{13},k))}^{t}c_{1,b(g_{13},k)}^{y_{i}}\\
+ \sum_{z_{i}\in Z} \sigma_{z_{i}}^{t} \lambda_{a_{z_{i},1},f(a_{z_{i},1},b(g_{13},k))}^{t}c_{1,b(g_{13},k)}^{z_{i}} = 0, \quad t\in[0,r-1],k\in[0,2].
\IEEEyesnumber\\
\end{IEEEeqnarray*}

Applying Claim \ref{cl:propf} with $i = a_{i,1}, i_1 = a_{1,1}=a_{2,1}, i_2 = a_{3,1}$ and using $g_{13}$ instead of $g_{12}$, we can define the following:
 \begin{IEEEeqnarray*}{l}
 \label{eqn19}
 \lambda_{a_{i,1}}:= \lambda_{a_{i,1},f(a_{i,1},b(g_{13},0))} =  \lambda_{a_{i,1},f(a_{i,1},b(g_{13},1))} =  \lambda_{a_{i,1},f(a_{i,1},b(g_{13},2))} , a_{i,1} \not \in \{a_{1,1},a_{3,1}\}
 \IEEEyesnumber\\
 \label{eqn20}
 \lambda_{a_{1,1},0}^{'}:= \lambda_{a_{1,1},f(a_{1,1},b(g_{13},0))} =  \lambda_{a_{1,1},f(a_{1,1},b(g_{13},2))}, \qquad
\lambda_{a_{1,1},1}^{'}:= \lambda_{a_{1,1},f(a_{1,1},b(g_{13},1))},
\IEEEyesnumber\\
\label{eqn21}
\lambda_{a_{3,1},0}^{'}:= \lambda_{a_{3,1},f(a_{3,1},b(g_{13},0))} =  \lambda_{a_{3,1},f(a_{3,1},b(g_{13},1))}, \qquad
\lambda_{a_{3,1},1}^{'}:= \lambda_{a_{3,1},f(a_{3,1},b(g_{13},2))}.\IEEEyesnumber 
\end{IEEEeqnarray*}
${\lambda_{a_{1,1},0}^{'},\lambda_{a_{1,1},1}^{'},\lambda_{a_{3,1},0}^{'},\lambda_{a_{3,1},1}^{'},\lambda_{a_{i,1}}},$
$a_{i,1} \not \in \{a_{1,1},a_{3,1}\}$ are all different.

Using the notation defined in (\ref{eqn19})-(\ref{eqn21}), we can write (\ref{eqn18}) for $k \in [0,1]$ 
 and sum over $k \in [0,1]$ resulting in
\begin{IEEEeqnarray*}{l}
\label{eqn22}
\sigma_{1}^{t}\sum_{k=0}^{1} (\lambda_{a_{1,1},k}^{'})^{t}c_{1,b(g_{13},k)}^{1}
+ \sigma_{2}^{t}\sum_{k=0}^{1} (\lambda_{a_{1,1},k}^{'})^{t}c_{1,b(g_{13},k)}^{2}
+ \sigma_{3}^{t} (\lambda_{a_{3,1},0}^{'})^{t}\sum_{k=0}^{1}c_{1,b(g_{13},k)}^{3}\\
+\sum_{w_{i} \in W} \sigma_{w_{i}}^{t}\sum_{k=0}^{1} (\lambda_{a_{1,1},k}^{'})^{t}c_{1,b(g_{13},k)}^{w_{i}} 
+ \sum_{y_{i}\in Y} \sigma_{y_{i}}^{t} (\lambda_{a_{3,1},0}^{'})^{t}\sum_{k=0}^{1}c_{1,b(g_{13},k)}^{y_{i}} \\
+ \sum_{z_{i}\in Z} \sigma_{z_{i}}^{t} (\lambda_{a_{z_{i},1}})^{t}\sum_{k=0}^{1}c_{1,b(g_{13},k)}^{z_{i}} = 0 , t \in [0,r-1].
\IEEEyesnumber\\
\end{IEEEeqnarray*}
Let 
\begin{IEEEeqnarray*}{l}
\mu_{3,1,1}^{(b)} = \sum_{k=0}^{1} c^{3}_{1,b(g_{13},k)}, \quad
\mu_{y_{i},1,1}^{(b)} = \sum_{k=0}^{1} c^{y_{i}}_{1,b(g_{13},k)}, \quad
\mu_{z_{i},1,1}^{(b)} = \sum_{k=0}^{1} c^{z_{i}}_{1,b(g_{13},k)}. 
\end{IEEEeqnarray*}
\eqref{eqn22} can be rewritten as
\begin{IEEEeqnarray*}{l}
\label{eqn23}
L_{1}+L_{2}+L_{3}+L_{4}+L_{5}+L_{6}=0, \IEEEyesnumber
\textnormal{where}
\end{IEEEeqnarray*}
\begin{IEEEeqnarray*}{l}
L_{j}=E_{L_{j}}F_{L_{j}} ,j \in \{1,3,4\}, \: 
L_{2}=\sum_{w_{i}\in W} E_{w_{i}}F_{w_{i}}, \\
L_{5}=E_{Y}F_{Y} , L_{6}=E_{Z}F_{Z} .\: \textnormal{Particularly}\\
E_{L_{1}}=[\sigma_{1}^{t}(\lambda_{a_{1,1},k}^{'})^{t}]_{t \in [0,r-1],k \in [0,1]}, \qquad
F_{L_{1}}=[c_{1,b(g_{13},k)}^{1}]_{k \in [0,1]},\\
E_{w_{i}}=[\sigma_{w_{i}}^{t}(\lambda_{a_{1,1},k}^{'})^{t}]_{t \in [0,r-1],k \in [0,1]}, \qquad
F_{w_{i}}=[c_{1,b(g_{13},k)}^{w_{i}}]_{k \in [0,1]},\\
E_{L_{3}}=[\sigma_{2}^{t}(\lambda_{a_{1,1},k}^{'})^{t}]_{t \in [0,r-1],k \in [0,1]},\qquad
F_{L_{3}}=[c_{1,b(g_{13},k)}^{2}]_{k \in [0,1]},\\
E_{L_{4}}=[\sigma_{3}^{t}(\lambda_{a_{3,1},0}^{'})^{t}]_{t \in [0,r-1]},\qquad\qquad\:\:\:
F_{L_{4}}=\mu_{3,1,1}^{(b)},\\
E_{Y}=[\sigma_{y_{i}}^{t}(\lambda_{a_{3,1},0}^{'})^{t}]_{t \in [0,r-1],i \in [1,|Y|]},\quad\:\:\:
F_{Y}=[\mu_{y_{i},1,1}^{(b)}]_{i \in [1,|Y|]},\\
E_{Z}=[\sigma_{z_{i}}^{t}(\lambda_{a_{z_{i},1}})^{t}]_{t \in [0,r-1],i \in [1,|Z|]}, \qquad
F_{Z}=[\mu_{z_{i},1,1}^{(b)}]_{i \in [1,|Z|]}.
\end{IEEEeqnarray*}
We now construct a matrix $P_{2}$ and on left-multiplying (\ref{eqn23}) with it, we get
\begin{IEEEeqnarray*}{l}
P_{2}L_{1}+P_{2}L_{2}+P_{2}L_{3}+P_{2}L_{4}+P_{2}L_{5}+P_{2}L_{6}=0. \\
p_0(x)=\Pi_{k=0}^{1}(x-\sigma_{1}\lambda_{a_{1,1},k}^{'})(x-\sigma_{2}\lambda_{a_{1,1},k}^{'}) \quad
\textnormal{and} \\
p_{i}(x)=x^{i}p_{0}(x) \quad \textnormal{for} \: i=0,1,..r-5. \\ \qquad
\end{IEEEeqnarray*}
For all $i \in [0,r-5]$, the degree of $p_{i}(x) < r$, hence
\begin{IEEEeqnarray*}{l}
p_{i}(x)=\sum_{j=0}^{r-1}p_{ij}x^{j} , \qquad i=0,1,..r-5. 
\end{IEEEeqnarray*}
The $(r-4)\times r$ matrix $P_{2}$ is defined as 
 \[P_{2} = \begin{bmatrix}
p_{0,0}&p_{0,1}&\cdots &p_{0,r-1}\\
p_{1,0}&p_{1,1}&\cdots &p_{1,r-1}\\
\vdots & \vdots & \vdots & \vdots\\
p_{r-5,0}&p_{r-5,1}&\cdots &p_{r-5,r-1}\\
\end{bmatrix}.\]
\\

For repair, we give a brief description of the steps to be performed at the first replacement node (the procedure is similar to that of the case of $a_{1,1} \not = a_{2,1}$), 
\begin{itemize}
    \item $P_{2}(L_{1}+L_{3})=0$ as $P_{2}[E_{L_{1}}\quad E_{L_{3}}]=0$.
    \item We can compute $P_{2}L_{2},P_{2}L_{5}$ at the first replacement node by downloading the symbols $\{c^{w_{i}}_{1,b(g_{13},0)},c^{w_{i}}_{1,b(g_{13},1)}\}$ of $F_{w_{i}}, \forall w_{i} \in W,\{\mu_{y_{i},1,1}^{(b)}\}$ of $F_{Y}, \forall y_{i} \in Y$ respectively. Then, we have
    $P_{2}(L_{4}+L_{6})=-P_{2}L_{2}-P_{2}L_{5}$.
    \item We can compute the remaining symbols of $\{F_{L_{4}}\cup F_{Z}\}$ at the first replacement node by downloading the symbols $\{c^{w_{i}}_{1,b(g_{13},0)},c^{w_{i}}_{1,b(g_{13},1)}\}$ of $F_{w_{i}}, \forall w_{i} \in W$,
     $\{\mu_{y_{i},1,1}^{(b)}\}$ symbols of $F_{Y}, \forall y_{i} \in Y$,\:any subset of symbols of size $k^{''}=(|Z|-(r-5))$ from $\{F_{L_{4}}\cup F_{Z}\}$ ($|Z|-(r-5) > 0 $ since $q > r$ and $|Z| \geq q-2$), from  $P_{2}(L_{4}+L_{6})=-P_{2}L_{2}-P_{2}L_{5}$ by inverting a square submatrix of $P_{2}[E_{L_{4}}\quad E_{Z}]$. Next, $F_{L_{1}}$ and $F_{L_{3}}$ can be recovered from
    \begin{IEEEeqnarray*}{l}
    L_{1}+L_{3}=-L_{2}-L_{5}-(L_{4}+L_{6})
    \end{IEEEeqnarray*}
    by inverting a square submatrix of $[E_{L_{1}}\quad E_{L_{3}}]$.\\
\end{itemize}

Similarly using the notation defined in (\ref{eqn19})-(\ref{eqn21}), we can write (\ref{eqn18}) for $k=2$ as follows:
\begin{IEEEeqnarray*}{l}
\label{eqn24}
L_{1}+L_{2}+L_{3}+L_{4}+L_{5}+L_{6}=0, \IEEEyesnumber
\textnormal{where}
\end{IEEEeqnarray*}
\begin{IEEEeqnarray*}{l}
L_{j}=E_{L_{j}}F_{L_{j}} ,j \in \{1,3,4\}, \: 
L_{2}=E_{W}F_{W},\\
 L_{5}=E_{Y}F_{Y} , L_{6}=E_{Z}F_{Z} .\: \textnormal{Particularly}\\
E_{L_{1}}=[\sigma_{1}^{t}(\lambda_{a_{1,1},0}^{'})^{t}]_{t \in [0,r-1]},\qquad\qquad
F_{L_{1}}=c_{1,b(g_{13},2)}^{1},\\
E_{W}=[\sigma_{w_{i}}^{t}(\lambda_{a_{1,1},0}^{'})^{t}]_{t \in [0,r-1],i \in [1,|W|]}, \:\:\:
F_{W}=[c_{1,b(g_{13},2)}^{w_{i}}]_{i \in [1,|W|]},\\
E_{L_{3}}=[\sigma_{2}^{t}(\lambda_{a_{1,1},0}^{'})^{t}]_{t \in [0,r-1]}, \qquad\qquad
F_{L_{3}}=c_{1,b(g_{13},2)}^{2},\\
E_{L_{4}}=[\sigma_{3}^{t}(\lambda_{a_{3,1},1}^{'})^{t}]_{t \in [0,r-1]}, \qquad\qquad
F_{L_{4}}=c_{1,b(g_{13},2)}^{3},\\
E_{Y}=[\sigma_{y_{i}}^{t}(\lambda_{a_{3,1},1}^{'})^{t}]_{t \in [0,r-1],i \in [1,|Y|]},\quad
F_{Y}=[c_{1,b(g_{13},2)}^{y_{i}}]_{i \in [1,|Y|]},\\
E_{Z}=[\sigma_{z_{i}}^{t}(\lambda_{a_{z_{i}},1})^{t}]_{t \in [0,r-1],i \in [1,|Z|]},\quad\:\:
F_{Z}=[c_{1,b(g_{13},2)}^{z_{i}}]_{i \in [1,|Z|]}.
\end{IEEEeqnarray*}
We now construct a matrix $P_{3}$ and on left-multiplying (\ref{eqn24}) with it, we get
\begin{IEEEeqnarray*}{l}
P_{3}L_{1}+P_{3}L_{2}+P_{3}L_{3}+P_{3}L_{4}+P_{3}L_{5}+P_{3}L_{6}=0. \\
p_0(x)=(x-\sigma_{1}\lambda_{a_{1,1},0}^{'})(x-\sigma_{2}\lambda_{a_{1,1},0}^{'}) \quad
\textnormal{and} \\
p_{i}(x)=x^{i}p_{0}(x) \quad \textnormal{for} \: i=0,1,..r-3. \\ \qquad
\end{IEEEeqnarray*}
For all $i \in [0,r-3]$, the degree of $p_{i}(x) < r$, hence
\begin{IEEEeqnarray*}{l}
p_{i}(x)=\sum_{j=0}^{r-1}p_{ij}x^{j} , \qquad i=0,1,..r-3.
\end{IEEEeqnarray*}
The $(r-2)\times r$ matrix $P_{3}$ is defined as 
 \[P_{3} = \begin{bmatrix}
p_{0,0}&p_{0,1}&\cdots &p_{0,r-1}\\
p_{1,0}&p_{1,1}&\cdots &p_{1,r-1}\\
\vdots & \vdots & \vdots & \vdots\\
p_{r-3,0}&p_{r-3,1}&\cdots &p_{r-3,r-1}\\
\end{bmatrix}.\]


For repair, we give a brief description of the steps to be performed at the second replacement node, 
\begin{itemize}
    \item $P_{3}(L_{1}+L_{3})=0$ as $P_{3}[E_{L_{1}}\quad E_{L_{3}}]=0$.
    \item We can compute $P_{3}L_{2},P_{3}L_{5}$ at the second replacement node by downloading the symbols $\{c^{w_{i}}_{1,b(g_{13},2)}\}$ of $F_{w_{i}}, \forall w_{i} \in W,\{c_{1,b(g_{13},2)}^{y_{i}}\}$ of $F_{Y}, \forall y_{i} \in Y$ respectively. Then, we have
    $P_{3}(L_{4}+L_{6})=-P_{3}L_{2}-P_{3}L_{5}$.
    \item We can compute the remaining symbols of $\{F_{L_{4}}\cup F_{Z}\}$ at the second replacement node by downloading the symbols $\{c^{w_{i}}_{1,b(g_{13},2)}\}$ of $F_{w_{i}}, \forall w_{i} \in W$,
     $\{c_{1,b(g_{13},2)}^{y_{i}}\}$ symbols of $F_{Y}, \forall y_{i} \in Y$,\:any subset of symbols of size $k^{'''}=(|Z|-(r-3))$ from $\{F_{L_{4}}\cup F_{Z}\}$ ($|Z|-(r-3) > 0 $ since $q > r$ and $|Z| \geq q-2$), from  $P_{3}(L_{4}+L_{6})=-P_{3}L_{2}-P_{3}L_{5}$ by inverting a square submatrix of $P_{3}[E_{L_{4}}\quad E_{Z}]$. Next, $F_{L_{1}}$ and $F_{L_{3}}$ can be recovered from
    \begin{IEEEeqnarray*}{l}
    L_{1}+L_{3}=-L_{2}-L_{5}-(L_{4}+L_{6})
    \end{IEEEeqnarray*}
    by inverting a square submatrix of $[E_{L_{1}}\quad E_{L_{3}}]$.\\
\end{itemize}

Please refer to the table for the summary of the downloads and recovery in the two rounds. After both the rounds, first replacement node recovers
$\{c^{1}_{1,b(g_{13},k)}:b_{g_{13}}=0,k \in \{0,1,2\}\} 
= \{c_{1,b}^{1}:b \in \{0,1,..l-1\}\}$
and second replacement node recovers
$\{c^{2}_{1,b(g_{13},k)}:b_{g_{13}}=0,k \in \{0,1,2\}\} 
= \{c_{1,b}^{2}:b \in \{0,1,..l-1\}\}.$
Both $c^{1}_{1}$ and $c^{2}_{1}$ are recovered.

\begin{enumerate}
\item The repair bandwidth for the case $a_{1,1}=a_{2,1}$ is given by
\begin{IEEEeqnarray*}{l}
RB_{a_{1,1} = a_{2,1}} = k^{''}\frac{l}{3} + |W|\frac{2l}{3} + |Y|\frac{l}{3} + \frac{l}{3}\\ \qquad\qquad\qquad
+ k^{'''}\frac{l}{3} + |W|\frac{l}{3} + |Y|\frac{l}{3} +\frac{2l}{3}\\ \qquad\qquad\quad\:
= (k^{''}+k^{'''})\frac{l}{3} + |W|l+|Y|\frac{2l}{3} +l. \IEEEyesnumber \label{eq:a1eqa2}
\end{IEEEeqnarray*}

where $k^{''}=|Z|-(r-5)$, $k^{'''}=|Z|-(r-3)$ and $|Z|=M-\frac{2M}{q}$.
\item Note that we need to compulsorily contact all the nodes given by the sets $W,Y$.
\end{enumerate}

\begin{table*}[ht]
    \centering
    \begin{tabular}{|c|c|c|}
    \hline
          & $1^{st}$ replacement node  & $2^{nd}$ replacement node\\
         \hline
         R1:&Download: $\{\mu_{i,1}^{(b)}:b_{g_{12}}=0\}$ from $k^{'}$ of $|\Gamma|$ nodes&
         Download: $\{\mu_{i,2}^{(b)}:b_{g_{12}}=0\}$ from $k^{'}$ of $|\Gamma|$ nodes\\
         
        $a_{1,1} \not = a_{2,1}$ &$\{c^{q_{i}}_{1,b(g_{12},0)},c^{q_{i}}_{1,b(g_{12},1)}:b_{g_{12}}=0\}$,$\forall q_{i} \in Q$, &  $\{c^{v_{i}}_{1,b(g_{12},0)},c^{v_{i}}_{1,b(g_{12},2)}:b_{g_{12}}=0\}$,$\forall v_{i} \in V$, \\
        
        & $\{ \sum_{k=0}^1 c_{1,b(g_{12},k)}^{v_{i}}:b
         _{g_{12}}=0\}$,$\forall v_{i} \in V$. & $\{ \sum_{k=0,2} c_{1,b(g_{12},k)}^{q_{i}}:b_{g_{12}}=0\}$,$\forall q_{i} \in Q$.\\
         
         &Recovery: $\{c^{1}_{1,b(g_{12},0)}:b_{g_{12}}=0\}$ 
         &Recovery:$\{c^{2}_{1,b(g_{12},0)}:b_{g_{12}}=0\}$\\
         &$\cup \{c^{1}_{1,b(g_{12},1)}:b_{g_{12}}=0\} \cup \{\mu_{2,1}^{(b)}:b_{g_{12}}=0\}$. & $\cup \{c^{2}_{1,b(g_{12},2)}:b_{g_{12}}=0\} \cup \{\mu_{1,2}^{(b)}:b_{g_{12}}=0 \}$. \\

          \hline
         R2:$a_{1,1} \not = a_{2,1}$&Download:$\{\mu_{1,2}^{(b)}:b_{g_{12}}=0 \}.$ &Download: $\{\mu_{2,1}^{(b)}:b_{g_{12}}=0 \}$.\\
         
         \hline
         
         R1:&Download:$\{\alpha_{i,1}^{(b)}:b_{g_{13}}=0\}$ from $k^{''}$ of $|Z^{'}|$ nodes &Download:$\{c^{i}_{1,b{(g_{13},2)}}:b_{g_{13}}=0\}$ from $k^{'''}$ of $|Z^{'}|$,\\
         
         
          $a_{1,1} = a_{2,1}$ &$\{c^{w_{i}}_{1,b(g_{13},0)},c^{w_{i}}_{1,b(g_{13},1)}:b_{g_{13}}=0\}$ $\forall w_{i} \in W$,&$\{c_{1,b(g_{13},2)}^{w_{i}}:b_{g_{13}}=0\}$ $\forall w_{i} \in W$,\\
          
          &$\{\sum_{k=0}^{1} c^{y_{i}}_{1,b(g_{13},k)}:b_{g_{13}}=0\}$ $\forall y_{i} \in Y$. &   $\{c_{1,b(g_{13},2)}^{y_{i}}:b_{g_{13}}=0\}$ $\forall y_{i} \in Y$.\\

          &Recovery:$\{c^{1}_{1,b(g_{13},0)}:b_{g_{13}}=0\}$ &Recovery:$\{c^{1}_{1,b(g_{13},2)}:b_{g_{13}}=0\}$\\
          
          & $\cup \{c^{1}_{1,b(g_{13},1)}:b_{g_{13}}=0\} 
        \cup \{c^{2}_{1,b(g_{13},0)}:b_{g_{13}}=0\}$ & $\cup \{c^{2}_{1,b(g_{13},2)}:b_{g_{13}}=0\}$.\\
 
         &$\cup \{c^{2}_{1,b(g_{13},1)}:b_{g_{13}}=0\}$.&\\
         
         \hline
         R2:$a_{1,1} =a_{2,1}$&Download:         $\{c^{1}_{1,b(g_{13},2)}:b_{g_{13}}=0 \}$.&Download: $\{c^{2}_{1,b(g_{13},0)},c^{2}_{1,b(g_{13},1)}:b_{g_{13}}=0 \}$. \\
         
         \hline
         
    \end{tabular}
    \caption{summary of downloads and recovery in the two rounds of cooperative repair, where $\mu_{i,1}^{(b)} = c^{i}_{1,b(g_{12},0)} + c^{i}_{1,b(g_{12},1)}$, $\mu_{i,2}^{(b)} = c^{i}_{1,b(g_{12},0)} + c^{i}_{1,b(g_{12},2)}$, $\alpha_{i,1}^{(b)} = c^{i}_{1,b(g_{13},0)} + c^{i}_{1,b(g_{13},1)}$ and  $Z^{'}=Z\cup\{3\}$.}
 \label{table1}
\end{table*}

\section{Repair Bandwidth Analysis of $\epsilon$-MSCR Codes} \label{sec:emscr_rb}
In this section, we give the repair bandwidth for the repair method described in Section \ref{sec:emscr_detailed} and compare it with that of the optimal. We also characterize the sub-packetization level.

%
%
%

\begin{lem}
The number of nodes $P$ contacted for the repair of two erasures (described in Section \ref{sec:emscr_repair})  is atleast $(M-r)$.
\end{lem}
\begin{IEEEproof}
For the case of $a_{1,1} \not = a_{2,1}$, number of nodes contacted is
\begin{align*}
 &\geq |Q|+|V|+|\Gamma|-(r-3) \\
&\geq M-2-r+3\\
&\geq M-r.
\end{align*}
For $a_{1,1} = a_{2,1}$, number of nodes contacted is
\begin{align*}
&\geq |W|+|Y|+|Z|-(r-5) \\
&\geq M-3-r+5\\
&\geq M-r.
\end{align*}

\end{IEEEproof}
\begin{note}
It is not straight-forward to characterize the number of compulsory nodes in terms of the hamming weight of the individual codewords as in \cite{guruswami2018epsilon}. Hence, we leave it for future work.
\end{note}

\begin{thm}

The repair bandwidth for the method described in Section \ref{sec:emscr_detailed} is at most $(1+\epsilon)$ times away from the optimal repair bandwidth where $\epsilon \leq (\frac{r}{P+1})(\frac{1}{2}+(2-\delta)\frac{P}{3})-1$, $P$ is the total number of nodes contacted for the repair of the two failed nodes. \\
\end{thm}
\begin{IEEEproof}
Let $\mathcal{P}$ denote the set of contacted nodes for repair and let $|\mathcal{P}| = P$. For a given node $i$ and each $j \in [N]$, if we assume that $a_{i,j}$ is helping the repair process via one of the sets $(Q,V,\Gamma, W, Y,Z)$, then we have the following upper bound on the total repair bandwidth.

\begin{IEEEeqnarray*}{l}
\label{eqn26}
RB_{tot} \leq|\{j \in [1,N]:a_{1,j} \not = a_{2,j}\}|\frac{2l}{3}
+ \:|\{j \in [1,N]:a_{1,j} = a_{2,j}\}|l
+\sum_{i \in \mathcal{P}}(|\{j \in [1,N]:a_{1,j} \not = a_{2,j}, a_{i,j} = a_{1,j}\}|l \\ \qquad\qquad
+ \:|\{j \in [1,N]:a_{1,j} \not = a_{2,j}, a_{i,j} = a_{2,j}\}|l 
+ \:|\{j \in [1,N]:a_{1,j} \not = a_{2,j}, a_{i,j} \not = a_{1,j},  a_{i,j} \not = a_{2,j}\}|\frac{2l}{3}  \\ \qquad\qquad
+ \:|\{j \in [1,N]:a_{1,j} = a_{2,j}, a_{i,j} = a_{1,j}\}|\:l 
+ \:|\{j \in [1,N]:a_{1,j} = a_{2,j}, a_{i,j}  = a_{3,j}\}|\frac{2l}{3}\\ \qquad\qquad
+ \:|\{j \in [1,N]:a_{1,j} = a_{2,j},a_{i,j} \not = a_{1,j}, a_{i,j} \not = a_{3,j}\}|\frac{2l}{3})
\IEEEyesnumber
\end{IEEEeqnarray*}
where the first two terms correspond to the repair bandwidth in the second round for $a_{1,1} \neq a_{2,1}$ and $a_{1,1} =  a_{2,1}$ respectively.
We note that
\begin{IEEEeqnarray*}{l}
\label{eqn27}
\hspace{-1.6in}\{j \in [1,N]:a_{1,j} \not = a_{2,j}\}|\frac{2l}{3}
+ \:|\{j \in [1,N]:a_{1,j} = a_{2,j}\}|l\\
\leq \{j \in [1,N]:a_{1,j} \not = a_{2,j}\}|l
+ \:|\{j \in [1,N]:a_{1,j} = a_{2,j}\}|l\\
\leq Nl.
\IEEEyesnumber
\end{IEEEeqnarray*}
and also
\begin{IEEEeqnarray*}{l}
\label{eqn28}
\hspace{-0.75in}|\{j \in [1,N]:a_{1,j} \not = a_{2,j}, a_{i,j} \not = a_{1,j},  a_{i,j} \not = a_{2,j}\}|\\
\quad= N - |\{j \in [1,N]:a_{1,j} \not = a_{2,j}, a_{i,j} = a_{1,j}\}|
- \: |\{j \in [1,N]:a_{1,j} \not = a_{2,j}, a_{i,j} = a_{2,j}\}|\\
\qquad - \: |\{j \in [1,N]:a_{1,j} = a_{2,j}, a_{i,j} = a_{1,j}\}| -\: |\{j \in [1,N]:a_{1,j} = a_{2,j}, a_{i,j}  = a_{3,j}\}|\\
\qquad - \:|\{j \in [1,N]:a_{1,j} = a_{2,j},a_{i,j} \not = a_{1,j}, a_{i,j} \not = a_{3,j}\}|.
\IEEEyesnumber
\end{IEEEeqnarray*}

Substituting (\ref{eqn27}), (\ref{eqn28}) in (\ref{eqn26}), we have
\begin{IEEEeqnarray*}{l}
RB_{tot}\leq Nl + \sum_{i \in \mathcal{P}} (\frac{2Nl}{3} + |\{j \in [1,N]:a_{1,j} \not = a_{2,j}, a_{i,j} = a_{1,j}\}|\frac{l}{3} 
+ \: |\{j \in [1,N]:a_{1,j} \not = a_{2,j}, a_{i,j} = a_{2,j}\}|\frac{l}{3}  \\
\qquad\qquad+ \: |\{j \in [1,N]:a_{1,j} = a_{2,j}, a_{i,j} = a_{1,j}\}|\frac{l}{3}) 
\IEEEyesnumber
\end{IEEEeqnarray*}
\begin{IEEEeqnarray*}{l}
\hspace{-0.7in}RB_{tot}\leq Nl + \sum_{i \in \mathcal{P}} (\frac{2Nl}{3} + |\{j \in [1,N]: a_{i,j} = a_{1,j}\}|\frac{l}{3} 
+  |\{j \in [1,N]:a_{1,j} \not = a_{2,j}, a_{i,j} = a_{2,j}\}|\frac{l}{3})  
\IEEEyesnumber
\end{IEEEeqnarray*}
\begin{IEEEeqnarray*}{l}
\hspace{-2.1in}\leq Nl + \sum_{i \in \mathcal{P}} \: (\frac{2Nl}{3} + 2(N-D)(\frac{l}{3}))  \qquad\qquad\qquad\qquad\qquad \\
\hspace{-2.1in}\leq Nl + \sum_{i \in \mathcal{P}}\: (\frac{4Nl}{3} - \frac{2Dl}{3}).
\IEEEyesnumber\\
\end{IEEEeqnarray*}

From (\ref{eqn1}), the optimal repair bandwidth for co-operative repair is
\begin{IEEEeqnarray*}{l}
(RB)_{opt} = \frac{h(h+d-1)l}{h+d-k} \\
\quad\qquad\quad=\frac{h(h+P-1)L}{h+P-\frac{K_{S}}{Nl}}\\
\quad\qquad\quad=\frac{2(P+1)Nl}{2+P-(M-r)}.
\IEEEyesnumber
\end{IEEEeqnarray*}
as $h = 2, K_{S} = (M - r)Nl, d = P ,\:L = Nl$ for our code.
Since $P \leq  M-2$, we have
\begin{IEEEeqnarray*}{l}
\label{eqn33}
\frac{2(P+1)Nl}{P-(M-r)+2} \geq \frac{2(P+1)Nl}{M-2-(M-r)+2}.
\IEEEyesnumber
\end{IEEEeqnarray*}
To derive an upper bound on $\epsilon$, we consider the following equations:
\begin{IEEEeqnarray*}{l}
\label{eqn34}
Nl + P(\frac{4Nl}{3} - \frac{2Dl}{3}) = (1+\epsilon)(\frac{2(P+1)Nl}{P-(M-r)+2}) 
=(1+\epsilon_{2})(\frac{2(P+1)Nl}{M-2 -(M-r)+2}).
\IEEEyesnumber
\end{IEEEeqnarray*}
From (\ref{eqn33}) and (\ref{eqn34}), $\epsilon \leq \epsilon_{2}$ . Also, considering first and third terms from (\ref{eqn34}), we have
\begin{IEEEeqnarray*}{l}
Nl+P(\frac{4Nl}{3} - \frac{2Dl}{3}) = (1+\epsilon_{2})(\frac{2(P+1)Nl}{r}) \\
\implies \frac{1}{2} + P(\frac{2}{3}-\frac{\delta}{3}) = \frac{(1+\epsilon_{2})(P+1)}{r}\\
\implies \epsilon_{2}= \frac{r}{P+1}(\frac{1}{2}+(2-\delta)\frac{P}{3})-1.
\end{IEEEeqnarray*}

Since $\epsilon \leq \epsilon_{2}$, we have
\begin{IEEEeqnarray*}{l}
\epsilon  \leq (\frac{r}{P+1})(\frac{1}{2}+(2-\delta)\frac{P}{3})-1.
\IEEEyesnumber
\end{IEEEeqnarray*}
Hence, the repair bandwidth of the repair method that we described in Section \ref{sec:emscr_detailed} is atmost $(1+\epsilon) = (\frac{r}{P+1})(\frac{1}{2}+(2-\delta)\frac{P}{3})$ times away from the optimal repair bandwidth completing the proof.

\end{IEEEproof}

\begin{cor}
For the case of $r=5$, our construction with the given repair procedure results in an $\epsilon$-MSCR code whose repair bandwidth is $(1+\epsilon) = \frac{5}{6}\left (\frac{3+(2-\delta)2P}{P+1}\right )$($\sim \frac{5}{3}$, when $\delta$ is close to 1 and $P$ is large) times away from the optimal repair bandwidth.
\end{cor}

%
%

\begin{thm}
Given positive integers $r,q,u$ and an $\epsilon > 0$, there exists an $(M,K_{S}=(M-r)Nl,d_{min}=r+1,L=Nl)_{\mathbb{B}}$ $\epsilon$-MSCR code satisfying the $(\mathscr{\tau},\mathscr{\tau}^{'})$ repair property with sub-packetization scaling logarithmicaly with $M$ for constant $q,u$ and the required field size of the order of $M$ for constant $q$.
\end{thm}
\begin{IEEEproof}
An $\epsilon$-MSCR code with length $M=q^{K}=q^{ug}$ and $L=Nl=N3^{m}$, $m=\binom{n}{2}$ is obtained by combining a $(n=q,k=q-r,l=3^{m})_{\mathbb{B}}$MSCR code with a scalar linear code $\mathcal{C}^{S}=(N,M=q^{K},D=\delta N)_{q}$ with $K=ug$ and $\frac{g}{N} \approx \frac{1}{\sqrt{q}-1}$(the choice of these parameters are obtained from Theorem 3.3 in \cite{guruswami2018epsilon}).
\begin{IEEEeqnarray*}{l}
M=q^{ug}.
\end{IEEEeqnarray*}
By taking log on both sides and using  $g \approx \frac{N}{\sqrt{q}-1}$, we get
\begin{IEEEeqnarray*}{l}
\log M=\frac{uN}{\sqrt{q}-1}\log q.
\end{IEEEeqnarray*}
But $L=N3^{\binom{n}{2}}=N3^{\binom{q}{2}}$, so we get
\begin{IEEEeqnarray*}{l}
\log M=\frac{uL}{3^{\binom{q}{2}}(\sqrt{q}-1)}\log q.
\end{IEEEeqnarray*}
Since $q,u$ are constant,we have $L=O_{q,u}(\log M)$.

To have distinct $\{\lambda_{i,j}\}_{i \in [n=q], j \in [0,1]}$ for construction of MSCR code, we need a field size of at-least $2q$. Next for distinct scalars $\{\sigma_{i}\}_{i \in [M]}$ for construction of $\epsilon$-MSCR code, we need a field size of at-least $M=q^{K}$. So, overall for the construction of $\epsilon$-MSCR code, we need a field size of at-least $2q^{K}q+1$. For constant $q$, the required field size is $O_{q}(q^{K})=O_{q}(M)$ which proves the above theorem.   
\end{IEEEproof}

\bibliography{biblio2}
\bibliographystyle{ieeetr}

\end{document}